\pdfoutput=1
\documentclass[11pt,a4paper]{article}
\makeatletter
\def\input@path{{supporting_files/}}
\makeatother
\usepackage{float}
\usepackage{jcappub}
\usepackage[T1]{fontenc}
\usepackage{amsmath,amssymb,amsthm,mathtools}
\usepackage{graphicx}
\usepackage{booktabs}
\usepackage{orcidlink}
\newtheorem{proposition}{Proposition}

\newcommand{\mpl}{M_{\rm Pl}}
\newcommand{\fs}{t_{\rm scr}}
\newcommand{\fsde}{\rho_{\rm FSDE}}

\title{Fast Scrambling Cosmology}

\author[a,1]{Aliv Sahoo\orcidlink{0009-0008-4745-0902},\note{Corresponding author.}}

\author[b,c]{Fatima Abbasi\orcidlink{0009-0005-6546-410X},}
\author[b,c]{Jake McNaughton\orcidlink{0000-0002-7423-8113},}
\author[d,c,b]{and Niayesh Afshordi\orcidlink{0000-0002-9940-7040}}

\affiliation[a]{Department of Physics, Indian Institute of Science, Bangalore 560012, India}
\affiliation[b]{Perimeter Institute for Theoretical Physics,\\31 Caroline Street North, Waterloo, Ontario, N2L 2Y5, Canada}
\affiliation[c]{Department of Physics and Astronomy, University of Waterloo \\200 University Avenue West, Waterloo, ON, N2L 3G1, Canada}
\affiliation[d]{Waterloo Centre for Astrophysics, University of Waterloo \\
200 University Avenue West, Waterloo, ON N2L 3G1, Canada}

\emailAdd{alivsahoo@iisc.ac.in}
\emailAdd{fabbasi@pitp.ca}
\emailAdd{jmcnaughton@pitp.ca}
\emailAdd{nafshordi@pitp.ca}

\abstract{
Quantum gravity arguments such as trans-Planckian censorship suggest that de
Sitter space can only last for a time of order $H^{-1}\ln(\mpl/H)$ -- corresponding to the
fast-scrambling time of black hole and cosmological horizons. We ask whether
this clock leaves an imprint on cosmic expansion, via a minimal modification
of the Raychaudhuri equation that adds a term $-\Gamma H^2/\ln(\mpl^2/H^2)$
to $\dot H$, with $\Gamma={\cal O}(1)$. The slowly varying
$\sigma\equiv\Gamma/\ln(\mpl^2/H^2)\ll 1$ is nearly constant at late times.
De Sitter space is replaced by a slowly evolving vacuum (an exact
Lambert-$W$ solution), and for constant $\sigma$ the expansion rate is
found in closed form for arbitrary matter sources. The correction acts as
dark energy with two components: a vacuum mode with $w=-1+2\sigma/3$, and a term that integrates over the history of the
cosmic fluids, so that any change in the equation of state of a species
leaves a permanent residual; e.g., neutrinos becoming non-relativistic shift the dark energy
density by ${\cal O}(\sigma m_\nu^4)$. A covariant clock-field completion
reproduces this background, and a term that rescales the gravitational
constant governing cosmic expansion removes its short-wavelength instability
in the dark-energy era. Big Bang
nucleosynthesis constraints give a proxy bound $\Gamma=-3.7\pm6.1$ (68\% confidence), while the negative-$\Gamma$
branch qualitatively follows the $w_0w_a$CDM dark energy evolution inferred from DESI DR2 at
$z\gtrsim1$. The early-time tracking terms amount to a similar rescaling, for which
existing fits to cosmic microwave background (CMB), baryon acoustic
oscillation (BAO) and supernova data translate into the proxy estimate
$\Gamma\simeq-2.8\pm0.9$ ($\sigma\simeq-0.01$ today). These background-level results motivate a dedicated likelihood analysis of
structure formation in this theory.
}


\begin{document}
\maketitle
\flushbottom

\section{Introduction}
\label{sec:intro}

Recent Dark Energy Spectroscopic Instrument (DESI) analyses, especially
extended-dark-energy combinations with cosmic microwave background (CMB)
and supernova data, continue to show a preference for mild low-redshift
departures from the minimal $\Lambda$CDM background expansion
\cite{DESI:2025zgx,DESIExtendedDE2025}. Whether this points to new microphysics, modified
gravity, or a statistical fluctuation is not yet clear. At the same time,
quantum gravity arguments have repeatedly suggested that an exactly stable
de Sitter phase may be too naive a description of the late universe.
Examples include the swampland program, the trans-Planckian censorship
conjecture (TCC), recent discussions of the origin and fate of our universe,
and horizon-scale modifications of gravity
\cite{BedroyaVafa2020,BedroyaInflation2020,Lehnert2025,Vafa2025,Mathur2025,WenEtAl2024,Brandenberger2025}.

This work is motivated specifically by the TCC. In its cosmological
form, the conjecture states that no fluctuation that was once trans-Planckian
should be stretched beyond the Hubble radius in a theory consistent with
quantum gravity \cite{BedroyaVafa2020,BedroyaInflation2020}. For an
approximately de Sitter phase with Hubble scale $H$, this implies a finite
lifetime of order
\begin{equation}
    \Delta t_{\rm TCC} \sim H^{-1}\ln\!\left(\frac{\mpl}{H}\right),
\end{equation}
since a mode that has a Planckian wavelength at the start of the phase is
stretched to $e^{H\Delta t}\mpl^{-1}$ and must remain inside the Hubble radius
$H^{-1}$.
Up to order-one factors, this is the same logarithmic timescale that appears
in the fast-scrambling conjecture for black holes and horizons
\cite{HaydenPreskill2007,SekinoSusskind2008,OshitaAfshordi2023}. The scrambling
time is the time it takes for information deposited in a system to spread over
all of its degrees of freedom, so that it can no longer be recovered from any sufficiently
small subsystem. For ordinary systems, with local interactions, it grows as a power of the number of
degrees of freedom, whereas black holes are conjectured to be the fastest
scramblers in nature, with $\fs\sim T^{-1}\ln S$ in terms of their temperature
$T$ and entropy $S$~\cite{SekinoSusskind2008}. The argument applies to any
Rindler-like horizon: a perturbation falling onto the stretched horizon spreads
over a horizon-sized region in a time of order the inverse temperature times the
logarithm of the ratio of the horizon size to the stretched-horizon
thickness~\cite{Susskind:2011ap}. The same reasoning underlies the search for
gravitational wave echoes from black hole
mergers~\cite{Cardoso:2016rao,Abedi:2016hgu,OshitaAfshordi2023}: if the
horizon is replaced by Planckian structure, perturbations reflected from it
return with a delay of order the scrambling time, $r_s\ln(r_s/\ell_{\rm Pl})$,
so that quantum horizon physics would leave an imprint on the ringdown. Fast
scrambling cosmology applies the same logic to the cosmological horizon, with
the expansion history in place of the ringdown. Since de
Sitter temperature and entropy obey
\begin{equation}
    T_{\rm dS}=\frac{H}{2\pi},
    \qquad
    S_{\rm dS}\sim \frac{\mpl^2}{H^2},
\end{equation}
the de Sitter scrambling time scales as
\begin{equation}
    \fs \sim \frac{1}{T_{\rm dS}}\ln S_{\rm dS}
    \sim \frac{1}{H}\ln\!\left(\frac{\mpl^2}{H^2}\right).
\end{equation}
This suggests that the same microscopic physics responsible for fast
scrambling may leave an observable imprint on cosmological evolution.

Motivated by the connection between the de Sitter scrambling timescale and the trans-Planckian censorship conjecture, the aim of the paper is to  study a phenomenological modification of the cosmological background equations, in which the Hubble rate acquires a decay (or growth) rate equal to the inverse scrambling time times an ${\cal O}(1)$ coefficient $\Gamma$ (Sec.~\ref{sec:model}). We derive the vacuum solution and obtain analytic background solutions in the slowly varying-log approximation, in which the effective dark-energy density consists of a vacuum mode and a contribution sourced by the history of the cosmological fluids (Sec.~\ref{sec:analytic}). We show that the modified equation cannot be obtained from a restricted class of local minisuperspace actions, and present a scalar-clock construction that reproduces the  background equation. We derive illustrative proxy constraints on the scrambling parameter from Big Bang nucleosynthesis (BBN)~\cite{Yeh:2026pil,Alvey:2019ctk} and, through a mapping onto the ``cosmic glitch in gravity'' model~\cite{Wen:2023wes,Wen:2026xlj}, from CMB, baryon acoustic oscillation (BAO) and supernova data (Sec.~\ref{sec:sigma_bounds}). Furthermore, we compare the predicted late-time dark-energy evolution with the Chevallier--Polarski--Linder (CPL)~\cite{Chevallier:2000qy,Linder:2002et} posterior inferred from DESI Data Release 2 (DR2) data (Sec.~\ref{sec:DESIoverlay}). We also investigate how the relativistic-to-non-relativistic transition of neutrinos can generate a residual contribution to the late-time dark-energy density (Sec.~\ref{sec:neutrino}). These results provide a preliminary assessment of the model at the background level, while highlighting the need for a consistent perturbative treatment and a dedicated cosmological likelihood analysis (Sec.~\ref{sec:discussion}).

\section{Fast-scrambling modification of FLRW cosmology}
\label{sec:model}

We work in a spatially flat Friedmann--Lema\^{\i}tre--Robertson--Walker (FLRW) background,
\begin{equation}
    ds^2=-dt^2+a^2(t)\,d\vec{x}^{\,2},
\end{equation}
and use reduced Planck units with $8\pi G = 1$ and $\mpl=1$ unless otherwise
stated. The standard background equations then read
\begin{equation}
    3H^2=\rho,
    \qquad
    \dot H=-\frac{1}{2}\,(\rho+P),
    \qquad
    \dot\rho+3H(\rho+P)=0,
    \label{eq:standard}
\end{equation}
where $H\equiv \dot a/a$.

The key phenomenological assumption is that the horizon should inherit an
instability rate set by the inverse scrambling time,
\begin{equation}
    \gamma_{\rm FS}\sim \fs^{-1}\sim
    \frac{H}{\ln(H^{-2})}.
\end{equation}
Since $\dot H$ carries units of $H^2$, the simplest correction to the standard background equations compatible
with this scaling is
\begin{equation}
    \dot H=
    -\frac{1}{2}(\rho+P)
    -\Gamma\frac{H^2}{\ln(H^{-2})},
    \label{eq:modified}
\end{equation}
with $\Gamma$ being a dimensionless parameter controlling the strength of the
fast-scrambling correction. For $H<\mpl=1$ the logarithm is positive, so for $\Gamma>0$ the new term drives $H$ downward, replacing exact de Sitter expansion by a slow decay of $H$ over a time $\sim\ln(\mpl^2/H^2)/(\Gamma H)$, i.e.\ of order the scrambling time; for $\Gamma<0$, it drives $H$ slowly upward instead (Sec.~\ref{subsec:vacuum}).

It is useful to define the dimensionless effective scrambling parameter
$\sigma(t)\equiv \Gamma/\ln(H^{-2})$, so that equation~\eqref{eq:modified} can be written as
$\dot H=-\frac{1}{2}(\rho+P)-\sigma(t)H^2$.
At this stage, $\sigma(t)$ is generally time-dependent. However, the logarithm in \eqref{eq:modified} varies only slowly when $H$ changes over
cosmological timescales and hence we will later consider the limit in which $\sigma(t)$ can be approximated by a constant. Numerically, $\ln(\mpl^2/H^2)\simeq277$ today, $\simeq257$ at recombination and $\simeq197$ at nucleosynthesis, so that $\sigma$ changes by less than $10\%$ between recombination and today (but by $\sim40\%$ between nucleosynthesis and today), and $|\Gamma|={\cal O}(1)$ corresponds to $|\sigma|$ of a few times $10^{-3}$.

Equation~\eqref{eq:modified} should be interpreted as an effective, coarse-grained equation
governing the background cosmological evolution, rather than as a complete
modified-gravity theory. We assume that the ordinary matter components
continue to satisfy the standard continuity equation. We then retain the
standard FLRW Hamiltonian constraint, $3H^2=\rho_{\rm tot},$
and define the additional contribution induced by the fast-scrambling term as
an effective fast-scrambling dark-energy (FSDE) density, $\rho_{\rm FSDE}$, through $
3H^2=\rho+\rho_{\rm FSDE}.$

\section{Analytic solutions and action principle}
\label{sec:analytic}

\subsection{A no-go result for a class of minisuperspace actions}
\label{subsec:nogo}

We first answer the question of whether \eqref{eq:modified} can be derived from a
local minisuperspace action. We show that,
within the class of actions considered below, the answer is no. The restriction to unimodular time below is motivated by unimodular gravity, in which the vacuum energy decouples from the dynamics and the cosmological constant arises as an integration constant; this is the natural setting for a modification whose vacuum limit is itself nontrivial.

\begin{proposition}
Consider an FLRW minisuperspace action in unimodular time, i.e.\ with the
lapse fixed to $N(u)=a(u)^{-3}$ (equivalently $\det(-g)=1$, as in unimodular gravity), of the form
\begin{equation}\label{eq:action}
    S=\int du\, f(R,H),
\end{equation}
for a spatially flat metric. If the resulting equations of motion
remain regular in the vacuum limit $\rho+P\to 0$, then the action \eqref{eq:action} cannot reproduce
the modified Raychaudhuri equation \eqref{eq:modified}.
\end{proposition}

\begin{proof}
Write the metric in unimodular time $u$ as
\begin{equation}
    ds^2=-N^2(u)\,du^2+a^2(u)\,d\vec{x}^{\,2},
    \qquad
    N(u)=a^{-3}(u),
\end{equation}
and define
\begin{equation}
    x\equiv \frac{a^3}{3},
    \qquad
    \dot x = \frac{dx}{du}=H.
\end{equation}
In this gauge, the Ricci scalar is
\begin{equation}
    R=6\left(2H^2+a^3\frac{dH}{du}\right)
    =12\dot x^2+18x\ddot x.
\end{equation}
Therefore the action becomes
\begin{equation}
    S=\int du\, f\!\left(12\dot x^2+18x\ddot x,\dot x\right).
\end{equation}
The Euler-Lagrange equation for a Lagrangian depending on $x$, $\dot x$, and
$\ddot x$ is
\begin{equation}
    18\ddot x\,f_{(1,0)}
    -\frac{d}{du}\left(24\dot x\,f_{(1,0)}+f_{(0,1)}\right)
    +\frac{d^2}{du^2}\left(18x\,f_{(1,0)}\right)=0,
    \label{eq:euler-lagrange}
\end{equation}
where $f_{(m,n)}$ denotes derivatives of $f$ with respect to its first and
second arguments. If $\dot x=H$ is constant, then $\ddot x=0$ and
every term in \eqref{eq:euler-lagrange} vanishes identically. Hence every
action in this class necessarily admits a nontrivial constant-$H$ vacuum
solution.

By contrast, the vacuum limit of \eqref{eq:modified} is
\begin{equation}
    \dot H = -\Gamma \frac{H^2}{\ln(H^{-2})},
    \label{eq:vacuum}
\end{equation}
which has no nonzero constant solution when $\Gamma \ne0$. This contradiction
shows that the modified equation cannot follow from the assumed action
principle.
\end{proof}

The proposition is a statement about the reduced variational problem, in which the lapse is fixed and the Hamiltonian constraint is
replaced by an integration constant, as in unimodular gravity. It does not
exclude more elaborate completions involving additional degrees of freedom,
nonlocality, or explicitly dissipative/open system dynamics. It does, however, strongly suggest that
\eqref{eq:modified} should be treated as an effective description rather than
as a conventional local gravitational action.

\subsection{A covariant completion with a scalar clock}
\label{subsec:moregeneralactions}
The no-go result of the preceding subsection applies to the restricted
minisuperspace class $f(R,H)$. It does not, however, exclude more general
covariant actions. In particular, introducing a scalar clock allows us to
construct a covariant action that reproduces the constant-$\sigma$ limit of
the modified background equation. Define
\begin{equation}
 X\equiv\nabla_\mu\phi\nabla^\mu\phi,
 \qquad
 u_\mu\equiv-\frac{\nabla_\mu\phi}{\sqrt{-X}},
 \qquad
 \Theta\equiv\nabla_\mu u^\mu.
 \label{eq:clock-kinematics}
\end{equation}
  Consider the action
\begin{equation}
 S=\int d^4x\sqrt{-g}\left[\frac{\mpl^2}{2}R+F(X,\Theta,\phi)\right]
 +S_{\mathrm{other}},
 \label{eq:clock-action}
\end{equation}
where
\begin{equation}
 F(X,\Theta,\phi)=-V(\phi)
 +\frac{\sigma\mpl^2}{9}\Theta^2\ln(-X),
 \label{eq:clock-lagrangian}
\end{equation}
and  $S_{\mathrm{other}}$ denotes the action for all components other than dark energy (matter,
radiation and neutrinos). Here $\phi$ has dimensions of time, so that $-X$ is dimensionless and equals one for the clock solution $\phi=t$ in proper time; we set $\mpl=1$ in what follows. As in any unitary-gauge description of dark energy, the construction is tied to a chosen background: the scalar equation of motion, equivalently $\nabla_\mu T_D^{\mu\nu}=0$, is what fixes $V(\phi)$ for a given source history (see the second remark below).

For a general $F(X,\Theta,\phi)$, variation with respect to the inverse
metric $g^{\mu\nu}$ gives
\begin{align}
 T_{\mu\nu}={}&\left(2XF_{,X}+\dot F_{,\Theta}\right)u_\mu u_\nu
 +2u_{(\mu}\nabla_{\nu)}F_{,\Theta} \nonumber\\
 &+\left(F-\dot F_{,\Theta}-F_{,\Theta}\Theta\right)g_{\mu\nu},
 \label{eq:clock-stress-tensor}
\end{align}
where $F_{,X}\equiv\partial F/\partial X$,
$F_{,\Theta}\equiv\partial F/\partial\Theta$, and
a dot denotes the
derivative along the  flow, $\dot{A}\equiv u^\mu\nabla_\mu A$.  The energy density and isotropic
pressure in the local  frame  are therefore
\begin{align}
 \rho_{D}&= T_{\mu\nu}u^\mu u^\nu
       =2XF_{,X}-F+F_{,\Theta}\Theta, \label{eq:clock-density}\\
 p_{D}&=\frac13\Delta^{\mu\nu}T_{\mu\nu}
       =F-\dot F_{,\Theta}-F_{,\Theta}\Theta,
 \qquad \Delta_{\mu\nu}\equiv g_{\mu\nu}+u_\mu u_\nu.
 \label{eq:clock-pressure}
\end{align}
and hence
\begin{equation}
 \rho_{D}+p_{D}=2XF_{,X}-\dot F_{,\Theta}.
 \label{eq:EFT-Enthalpy}
\end{equation}

In a homogeneous FLRW background, we have the freedom to choose unitary gauge,
$\phi=t$. This gives $X=-1$, $\dot X=0$, and $\Theta=3H$.  For the  Lagrangian
\eqref{eq:clock-lagrangian}, we find:
\begin{equation}
 2XF_{,X}=\frac{2\sigma}{9}\Theta^2,
 \qquad
 F_{,\Theta}=\frac{2\sigma}{9}\Theta\ln(-X)=0,
 \qquad
 \dot F_{,\Theta}=0.
 \label{eq:clock-derivatives}
\end{equation}
Thus
\begin{equation}
 \rho_{D}+p_{D}=\frac{2\sigma}{9}\Theta^2=2\sigma H^2.
 \label{eq:clock-equation-of-state}
\end{equation}
Together with the Einstein Raychaudhuri equation, this gives
\begin{equation}
 \dot H=-\frac12(\rho+p)-\sigma H^2,
 \label{eq:clock-modified-raychaudhuri}
\end{equation}
which is precisely the constant-log fast-scrambling equation.

This construction establishes that the minisuperspace no-go result does not exclude
more general covariant actions with a preferred clock field.
Three remarks are in order. First, the construction requires a preferred time field: the clock $\phi$ spontaneously breaks Lorentz invariance, as in ghost-condensate, cuscuton or Ho\v{r}ava-type theories, and as in the cosmic-glitch models of refs.~\cite{Wen:2023wes,WenEtAl2024} to which we compare in Sec.~\ref{sec:sigma_bounds}. Second, as in any unitary-gauge description of dark energy, the potential is not a free function: varying the action with respect to the lapse and the scale factor in the gauge $\phi=t$ gives $3H^2=\rho+V+2\sigma H^2$ and $2\dot H+3H^2=-P+V$, so that $V(t)=(3-2\sigma)H^2-\rho(t)$ is fixed by the background (for a generic $V(\phi)$, e.g.\ a constant, $\phi=t$ is not a solution and the theory describes a different cosmology). The content of the construction is therefore the operator $\Theta^2\ln(-X)$, which generates $\rho_D+p_D=2\sigma H^2$ for any matter content. Third, the construction is not perturbatively viable. Expanding the action to quadratic order in unitary gauge (Appendix~\ref{app:perturbations}), the tensor sector is unmodified, while the scalar sector corresponds, in the language of the effective theory of dark energy~\cite{Gleyzes:2014rba,Bellini:2014fua}, to
\begin{equation}
 \alpha_K=-2\sigma,\qquad \alpha_B=\frac{4\sigma}{3},\qquad \alpha_M=\alpha_T=0 .
 \label{eq:alphas}
\end{equation}
The absence of a ghost requires $\alpha_K+\tfrac32\alpha_B^2=-2\sigma(1-\tfrac43\sigma)>0$, i.e.\ $\sigma<0$ for the small $|\sigma|$ of interest (the inequality also admits a remote branch $\sigma>3/4$), which is the sign indicated by the proxy estimates of Sec.~\ref{sec:sigma_bounds}, but the sound speed of the scalar mode,
\begin{equation}
 c_s^2=\frac{(\rho+P)/(3H^2)-\tfrac53\left(1-\tfrac23\sigma\right)}{1-\tfrac43\sigma},
 \label{eq:cs2}
\end{equation}
is negative at all epochs: $c_s^2\simeq-1/3$, $-2/3$ and $-5/3$ in radiation, matter and dark-energy domination, respectively, for $|\sigma|\ll1$. The construction therefore suffers from a gradient instability on all sub-horizon scales, and should be regarded as an existence proof at the background level rather than as a viable completion. Since the background fixes only $\rho_D+p_D=2\sigma H^2$, while the kineticity and braiding functions of the effective theory are free, the instability can be cured by additional operators; we comment on this in Sec.~\ref{sec:discussion}.

\subsection{Exact vacuum solution}
\label{subsec:vacuum}

Setting $\rho+P=0$ reduces the modified equation to \eqref{eq:vacuum}. This
equation can be solved exactly in terms of the Lambert-$W$ function. Writing
$t_0$ for an integration constant, one convenient form is
\begin{equation}
    H(t)=
    \frac{2\,W\!\left(\dfrac{\Gamma (t-t_0)}{2e}\right)}
    {\Gamma (t-t_0)}.
    \label{eq:lambert}
\end{equation}
For $0<H<\mpl$ the relevant solution is the principal branch $W_0$; the other
real branch, $W_{-1}$, describes the far side of the Planck-scale turning point
discussed below. Standard general relativity would have $H={\rm constant}$ in vacuum; here, for $\Gamma>0$, the Hubble rate
slowly decays. At late times,
\begin{equation}
    H(t)\sim \frac{2\ln[\Gamma (t-t_0)]}{\Gamma (t-t_0)},
    \qquad
    t\to \infty, \qquad (\Gamma>0),
    \label{eq:late}
\end{equation}
which makes the de Sitter instability explicit.
For $\Gamma<0$ the solution is the time reverse of the above, since equation~\eqref{eq:vacuum} is invariant under $t\to-t$, $\Gamma\to-\Gamma$: starting from $H\ll\mpl$, the Hubble rate slowly grows and reaches the Planck scale, $H=\mpl$, at $t=t_0+2/|\Gamma|$, where the two real branches of $W$ meet and the effective description ceases to apply. Starting from a Hubble rate $H$, this takes a time $\Delta t\simeq \ln(\mpl^2/H^2)/(|\Gamma|H)=\fs/|\Gamma|$. In the constant-$\sigma$ approximation of the next subsection, this appears as the familiar Big Rip pole, $H\simeq1/[|\sigma|(t_{\rm rip}-t)]$; the divergence is an artifact of holding $\sigma$ fixed, since the exact equation only reaches $H=\mpl$, but the timescale is reliable: for the value $\sigma_0\simeq-0.01$ suggested by the proxy estimates of Sec.~\ref{sec:sigma_bounds}, $t_{\rm rip}-t_{\rm now}\simeq100\,H_0^{-1}\simeq1.4$~Tyr. Either sign of $\Gamma$ therefore deforms the quasi-de Sitter phase on a timescale set by the scrambling time; they differ in whether $H$ decays or grows.

We stress, however, that neither branch realizes the TCC by itself. The conjecture is an integrated bound, $\int H\,dt<\ln(\mpl/H_f)$~\cite{BedroyaVafa2020}, whereas here the expansion remains accelerated. For the exact vacuum equation, the evolution of the Hubble parameter satisfies
\begin{align}
\ln^2\left(\frac{M_{\rm Pl}^2}{H^2}\right)
=
\ln^2\left(\frac{M_{\rm Pl}^2}{H_i^2}\right)+4\Gamma N, \\
\Rightarrow \frac{\int_{t_i}^{t_f} H\,dt}{\ln(\mpl/H_f)} = \frac{\ln^2\left(\frac{M_{\rm Pl}^2}{H_f^2}\right)-\ln^2\left(\frac{M_{\rm Pl}^2}{H_i^2}\right)}{2 \Gamma \ln\left(\frac{M_{\rm Pl}^2}{H_f^2}\right)}\label{eq:TCC_bound}
\end{align}
 where $N=\ln(a/a_i)= \int H dt$ is the number of e-folds, and $a_i$ and $H_i$ denote the scale factor and Hubble parameter at a reference epoch. It is then easy to see that the right-hand side of equation~\eqref{eq:TCC_bound} always exceeds unity for $\Gamma = {\cal O}(1)$ at late times if $H_i \ll M_{\rm Pl}$: the bound is saturated after $N\simeq \tfrac{1}{2}\ln(M_{\rm Pl}^2/H_i^2)\simeq 140$ e-folds for $|\Gamma|={\cal O}(1)$ and $\ln(M_{\rm Pl}^2/H_i^2)\simeq277$, while $H$ is still far below the Planck scale.
 The scrambling term thus sets the rate at which de Sitter space is deformed, on the timescale singled out by the TCC; satisfying the integrated bound would require an exit from accelerated expansion, or a finite domain of validity of the effective description, before that point.

\subsection{Constant-log approximation and the general solution}
\label{subsec:constant-log}

As discussed in Sec.~\ref{sec:model}, over a wide range of cosmic history,
$\sigma(t)\simeq\sigma$ is approximately constant. Using $ \dot H + \sigma H^2 = -\frac{1}{2}(\rho+P)$ and the continuity equation  one finds
\begin{equation}
    a\frac{dH^2}{da}+2\sigma H^2=\frac{1}{3}a\frac{d\rho}{da}.
\end{equation}
which has the solution
\begin{equation}
    H^2(a)=
    \frac{\rho(a)}{3}
    -\frac{2\sigma}{3}a^{-2\sigma}\int^a d\tilde a\,
    \tilde a^{\,2\sigma-1}\rho(\tilde a)
    + {\cal C}\,a^{-2\sigma},
    \label{eq:general-solution}
\end{equation}
where ${\cal C}$ is an integration constant. For pressureless matter and
radiation,
\begin{equation}
    \rho(a)=\rho_{m,0}a^{-3}+\rho_{r,0}a^{-4},
\end{equation}
the result collapses to
\begin{equation}
    H^2(a)=
    \frac{\rho_m(a)}{3-2\sigma}
    +\frac{\rho_r(a)}{3-\frac{3}{2}\sigma}
    +{\cal C}\,a^{-2\sigma}.
    \label{eq:mr}
\end{equation}
In other words, the fast-scrambling contributions in \eqref{eq:mr}, which track the matter and radiation densities with fixed fractional amplitudes (we refer to them as tracking terms), act as a renormalization of the gravitational coupling felt by the dominant fluid, $G_{\rm eff}/G=(1-\tfrac{2}{3}\sigma)^{-1}$ in matter domination and $(1-\tfrac12\sigma)^{-1}$ in radiation domination, while the vacuum mode takes the place of the cosmological constant. Tracking terms of this type are also found in the cosmic glitch model~\cite{Wen:2023wes,Wen:2026xlj,WenEtAl2024}, and they form the basis of the proxy constraints in Sec.~\ref{sec:sigma_bounds}.

Equation~\eqref{eq:general-solution} shows that the fast-scrambling
correction behaves like dynamical dark energy. Defining an effective density
through $3H^2=\rho+\fsde$, we obtain
\begin{equation}
    \fsde(a)=
    -2\sigma a^{-2\sigma}\int^a d\tilde a\,
    \tilde a^{\,2\sigma-1}\rho(\tilde a)
    + 3{\cal C}\,a^{-2\sigma}.
    \label{eq:fsde}
\end{equation}
The second term is the source-free solution of the linear equation for $H^2$
(its homogeneous solution, in the sense of ordinary differential equations); it
is the only piece that survives for $\rho=0$, where it coincides with the
constant-$\sigma$ limit of the vacuum solution of Sec.~\ref{subsec:vacuum}, and
we therefore refer to it as the vacuum mode. It reduces to a cosmological
constant when $\sigma\to 0$ and by itself has an effective equation of state
\begin{equation}
    w_{\rm vac}=-1+\frac{2\sigma}{3}.
\end{equation}
Thus a small negative $\sigma$ is mildly phantom-like, while a small positive
$\sigma$ is mildly quintessence-like. The first term in \eqref{eq:fsde} is
new: it remembers the full source history (we refer to it as the source-history integral) and can be driven by whichever
species dominate the integral.
 \section{Proxy constraints on $\sigma$ and $\Gamma$}

\label{sec:sigma_bounds}

In this section we obtain ballpark estimates of the fast-scrambling parameters $\sigma$ and $\Gamma$ by comparing the tracking terms of equation~\eqref{eq:mr} with existing constraints on a cosmological gravitational coupling that differs from Newton's constant. Following the cosmic glitch in gravity (CGG) model~\cite{Wen:2026xlj,Wen:2023wes}, let $G_{\rm cosmo}$ be the coupling that governs the cosmic expansion, with $\Omega_g \equiv 1-G_{\rm N}/G_{\rm cosmo}$. The Friedmann equation then becomes $3H^2=(\rho_{\rm nonDE}+\rho_\Lambda)/(1-\Omega_g)$, where $\rho_{\rm nonDE}$ contains matter, radiation, and neutrinos. This cosmic glitch in gravity can be interpreted as a dark energy component with density $\rho_{\rm DE}=\Omega_g\rho_{\rm nonDE}/(1-\Omega_g)+\rho_\Lambda/(1-\Omega_g)$. The first term tracks the dominant fluid with a fixed fractional amplitude $\Omega_g/(1-\Omega_g)$, in direct analogy with the tracking terms in our dark energy model~\eqref{eq:mr}. Comparing the  tracking coefficients in each era gives
\begin{equation}
\Omega_g=\tfrac{\sigma}{2}\ \ (\text{radiation domination}),\qquad
\Omega_g=\tfrac{2\sigma}{3}\ \ (\text{matter domination}).
\label{eq:cgg_map}
\end{equation}

In the matter era, we compare our results with the recent constraints on the CGG model~\cite{Wen:2026xlj}. These constraints, obtained using Planck, Atacama Cosmology Telescope (ACT) DR6, and South Pole Telescope (SPT)-3G D1 data combined with CMB lensing, DESI DR2 BAO, and supernova (SN) data from the Dark Energy Survey (DES, Dovekie compilation), yield $\Omega_g=-0.0073\pm0.0022$ (68\% confidence level, CL)~\cite{Wen:2026xlj}, which corresponds to $\sigma=-0.0110\pm0.0033$.   With $\sigma=\Gamma/\ln(M_{\rm Pl}^2/H^2)$ and
$\ln(M_{\rm Pl}^2/H^2)\simeq 257$ at recombination, we obtain $\Gamma\simeq-2.8\pm0.9$ (CMB+lensing+BAO+SN).

In the radiation era, BBN constrains  $G_{\rm cosmo}/G_{\rm N}$ at nucleosynthesis~\cite{Alvey:2019ctk}. The BBN$+\Omega_bh^2$ analysis of ref.~\cite{Alvey:2019ctk} gives $G_{\rm BBN}/G_{\rm N}=0.99^{+0.06}_{-0.05}$ at 95\% CL, which we approximate, by halving the interval, as $0.99^{+0.03}_{-0.025}$ at 68\% CL (a Gaussian approximation to an asymmetric 95\% interval). Since $G_{\rm BBN}/G_{\rm N}=1/(1-\Omega_g)$, this gives $\Omega_g=-0.010\pm0.028$ (68\% CL) and hence, by equation~\eqref{eq:cgg_map}, $\sigma=-0.020\pm0.055$ (68\% CL), which translates into $\Gamma\simeq-3.9\pm10.8$ using $\ln(\mpl^2/H^2)\simeq197$ at nucleosynthesis.

 We can also place another constraint on \(\sigma\) using  Standard Big Bang nucleosynthesis (SBBN) with a baseline of $N_\nu =3$. During the radiation-dominated era, the fast-scrambling modification effectively changes the expansion rate of the Universe, which can be interpreted as a shift in the number of relativistic degrees of freedom. To quantify this, we equate the standard expansion-rate modification,
$H^2/H_{\rm SBBN}^2 = 1 + (7/43)\Delta N_\nu,$
with the ratio
$G_{\rm BBN}/G_{\rm N} = \left(1-\sigma/2\right)^{-1}
$ of our model. Using recent BBN-only constraints from primordial helium and deuterium abundances~\cite{Yeh:2026pil}, we obtain
$\Delta N_\nu =-0.059\pm0.092$ and hence $\sigma=-0.019 \pm0.031$ at 68\% confidence, which corresponds
to  $\Gamma\simeq-3.7\pm6.1$.

\begin{figure}
    \centering
    \includegraphics[width=1\linewidth]{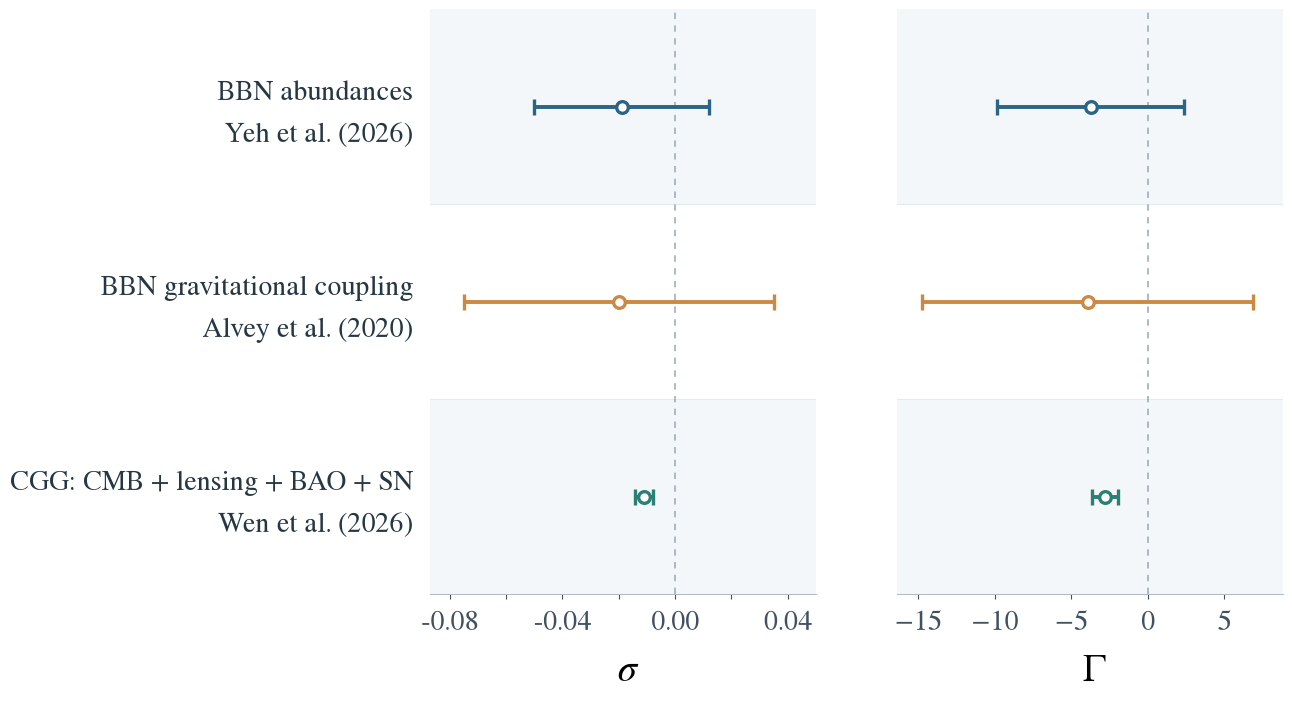}
    \caption[Illustrative constraints on $\sigma$ and $\Gamma$]{Illustrative constraints on the fast-scrambling parameters
$\sigma$ (left) and $\Gamma$ (right), inferred from
BBN light-element abundances~\cite{Yeh:2026pil},
BBN bounds on the gravitational coupling
~\cite{Alvey:2019ctk}, and the cosmic glitch in gravity (CGG)
analysis~\cite{Wen:2026xlj}.
Points denote central values (the means of the Gaussian-approximated translated constraints), while horizontal bars
show the corresponding 68\% confidence intervals.
The conversion between $\sigma$ and $\Gamma$ uses
$\ln(M_{\rm Pl}^2/H^2)\simeq197$ at nucleosynthesis
and $\simeq257$ at recombination.}
    \label{fig:constraints}
\end{figure}

Figure~\ref{fig:constraints} summarizes the three proxy estimates. All have negative central values and all are consistent with the ${\cal O}(1)$ magnitude of $\Gamma$ anticipated in Sec.~\ref{sec:model}; the two BBN-based estimates, which share the light-element inputs and are therefore not independent, are consistent with $\Gamma=0$, while the CGG-based proxy translates into $\Gamma\simeq-3$ at the $3\sigma$ level.

 We stress that these translated values are illustrative: the CMB and BAO constraints on the CGG model rely on its linear perturbations, which need not coincide with those of the fast-scrambling model, and the BBN bounds assume that the only effect at nucleosynthesis is the modified expansion rate.

\section{Late-time dark energy in light of DESI DR2}
\label{sec:DESIoverlay}
\begin{figure}[htbp]
    \centering
    \includegraphics[width=0.78\textwidth]{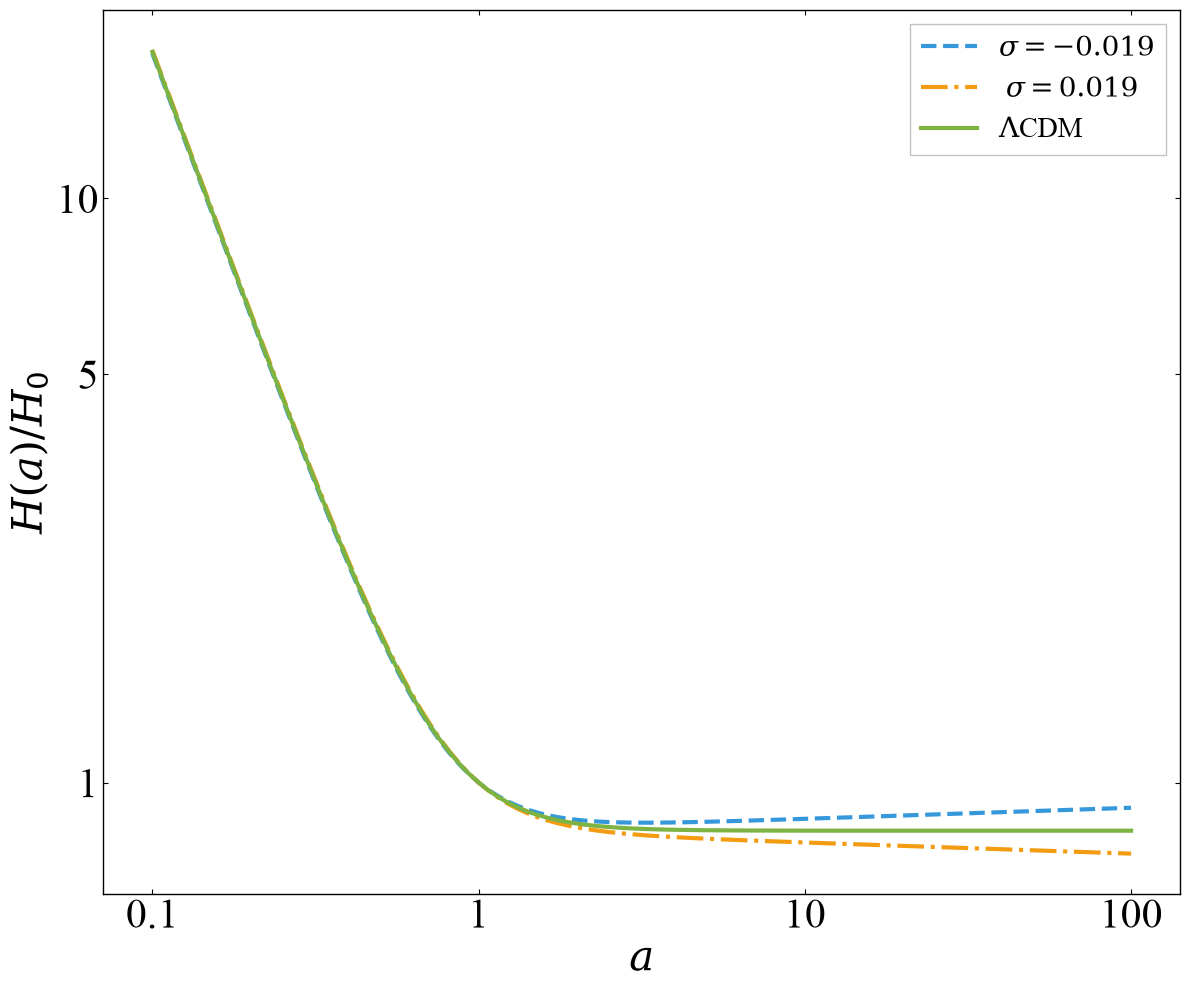}
    \caption[Normalized Hubble parameter for $\sigma=\pm0.019$]{Evolution of the normalized Hubble parameter, \(H(a)/H_0\), as a function of the scale factor \(a\) for the fast-scrambling model with \(\sigma = \pm 0.019\), compared with the standard \(\Lambda\)CDM prediction. 
    We adopt  $H_0 = 67.4\,\mathrm{km\,s^{-1}\,Mpc^{-1}}$, $\Omega_{m,0} = 0.315$, $\Omega_{\gamma,0} = 5.38 \times 10^{-5}$, and $\Omega_{\mathrm{DE,0}} \approx 0.683$ (fixed by flatness), with three  neutrino species in the normal hierarchy ($m_\nu = 0.05,  0.0086, 0.0\,\mathrm{eV}$), whose energy densities are evaluated using the Fermi--Dirac distribution. The present-day boundary condition is fixed by $\rho_{\mathrm{FSDE}}(a=1) = \Omega_{\mathrm{DE,0}}\rho_{c,0}$. The curves remain nearly indistinguishable in the past but separate noticeably in the future ($a > 1$).}
    \label{fig:hubble}
\end{figure}

We examine how the fast-scrambling modification affects the background expansion. 
Figure~\ref{fig:hubble}  illustrates the evolution of the normalized Hubble parameter, $H(a)/H_0$, for  scrambling parameter values of $\sigma=\pm 0.019$, chosen to reflect the central BBN value derived in Sec.~\ref{sec:sigma_bounds}, against a standard $\Lambda$CDM baseline. In the past ($a<1$) the three curves are nearly indistinguishable, but they separate from $\Lambda$CDM in the future ($a>1$), once the $a^{-2\sigma}$ mode dominates. The $\sigma<0$ branch
produces a larger late-time value of $H(a)$, while the $\sigma>0$ branch
leads to a smaller one, consistent with the phantom-like (quintessence-like) equation of state $w_{\rm vac}=-1+2\sigma/3$ of the vacuum mode and, for $\sigma<0$, with the slow approach to the Big Rip discussed in Sec.~\ref{subsec:vacuum}.

\begin{figure}[htbp]
    \centering
    \includegraphics[width=0.78\textwidth]{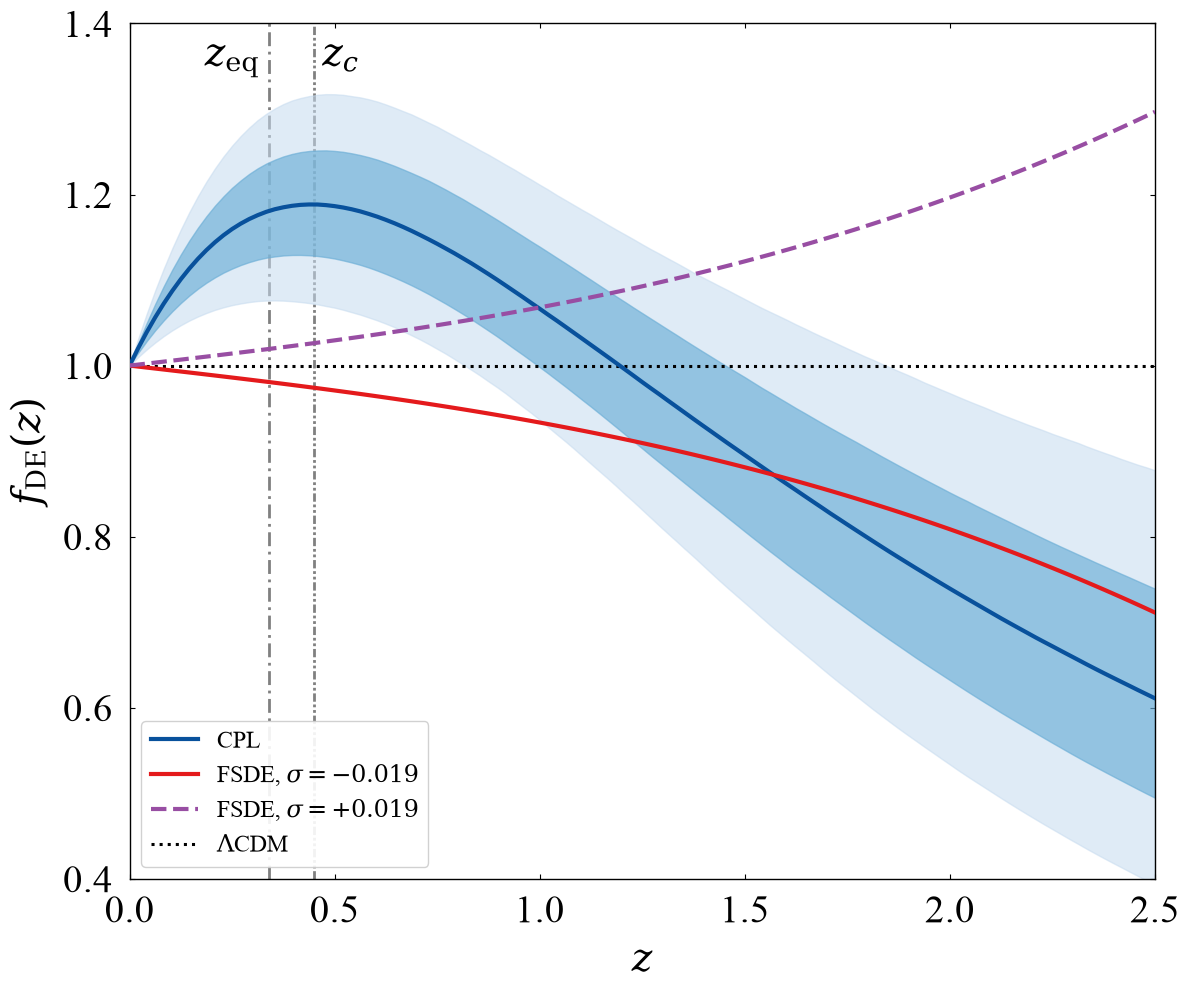}
    \caption[Fast-scrambling dark energy versus the DESI DR2 CPL posterior]{
Comparison of the normalized fast-scrambling dark-energy density, $f_{\rm DE}(z)\equiv\rho_{\rm FSDE}(z)/\rho_{\rm FSDE}(0)$, with the CPL dark-energy density inferred from the DESI DR2 BAO + Planck CMB + ACT DR6 lensing + Union3 data combination \cite{DESI:2025zgx}.
The solid blue curve shows the median of $f_{\rm DE}(z)$, calculated at each redshift using the sample weights supplied in the publicly available DESI DR2 chains. The darker and lighter shaded regions span the 16th--84th and 2.5th--97.5th percentiles, corresponding to 68\% and 95\% credible intervals, respectively.
 The red and purple curves show the fast-scrambling predictions for $\sigma=-0.019$ and $\sigma=+0.019$, respectively. The dotted horizontal line indicates the $\Lambda$CDM prediction,  and the vertical lines mark the dark-energy–matter equality redshift ($z_{\rm eq}$) and the phantom-crossing redshift ($z_c$), both for the CPL posterior median.  }
    \label{fig:de_desi}
\end{figure}

We next compare the evolution of the effective fast-scrambling dark energy with the CPL model of late-time dark energy \cite{DESI:2025zgx}. Figure~\ref{fig:de_desi} shows the normalized dark-energy density, $f_{\rm DE}(z) \equiv \rho_{\rm FSDE}(z)/\rho_{\rm FSDE}(0)$, for $\sigma=\pm0.019$, together with the posterior-median $w_0w_a$CDM trajectory and its $68\%$ and $95\%$ posterior bands, obtained from the publicly available DESI DR2 cosmological chains for the combination of DESI DR2 BAO, Planck CMB, ACT DR6 lensing, and Union3 supernova data.

The $\sigma=-0.019$ branch lies within the $68\%$ ($95\%$) CPL band for $z\gtrsim1.2$ ($z\gtrsim0.8$) and follows the posterior-median trajectory closely at $z\gtrsim1$. At lower redshift the CPL median exhibits a bump, $f_{\rm DE}\simeq1.2$ at $z\simeq0.45$, associated with its phantom crossing, which the fast-scrambling density does not reproduce. This is a firm prediction of the model at the background level: since $\dot\rho_{\rm FSDE}=-6\sigma H^3$ (Sec.~\ref{sec:discussion}), the effective dark energy density is monotonic in time for fixed-sign $\sigma$ and, wherever it is positive, its equation of state $w_{\rm FSDE}=-1+2\sigma H^2/\rho_{\rm FSDE}$ cannot cross $-1$. The same is true of the CGG model, which nevertheless fits CMB+BAO data as well as $w_0w_a$CDM with one parameter fewer (equal Bayesian evidence), and remains competitive when CMB lensing and supernovae are added~\cite{Wen:2026xlj}; this raises the possibility that the low-redshift feature is a property of the CPL parametrization rather than a requirement of the data, although only a likelihood analysis of the present model can resolve this.
Equivalently, this branch corresponds to a dark-energy density that
increases toward low redshift, producing the smoothly evolving,
phantom-like behavior that characterizes the DESI $w_0w_a$CDM reconstruction at $z\gtrsim z_c$.

 The result therefore suggests that the fast-scrambling
framework can reproduce, at the background level, the type of
smoothly varying late-time dark-energy behavior probed by the CPL model of the DESI
analysis~\cite{DESI:2025zgx}.
We stress that this comparison is not a likelihood analysis of the
fast-scrambling model.  A dedicated
constraint on $\sigma$ would require constructing the corresponding
fast-scrambling likelihood and consistently marginalizing over the
cosmological parameters and relevant observational datasets.

\section{Neutrino-sourced dark energy}
\label{sec:neutrino}

The fast-scrambling modification provides a simple mechanism through which the cosmic neutrino background  contributes to the late-time dark energy.
This mechanism is phenomenologically appealing because neutrinos naturally interpolate from a radiation-like phase to a matter-like phase at late times. In terms of comoving momentum $p$ and present-day neutrino temperature $T_{\nu,0}$, the energy density of a neutrino species with mass $m_\nu$ is given, for an ideal gas that decoupled instantaneously while relativistic, by the Fermi--Dirac distribution:
\begin{equation}
    \rho_\nu(a)=
    \frac{g_\nu}{2\pi^2 a^4}
    \int_0^\infty dp\,
    \frac{p^2\sqrt{p^2+m_\nu^2 a^2}}{e^{p/T_{\nu,0}}+1}.
    \label{eq:neutrino}
\end{equation}
with $T_{\nu,0}=(4/11)^{1/3}T_{\gamma,0}\simeq1.68\times10^{-4}$~eV and $g_\nu=2$ per species; the small spectral distortions from non-instantaneous decoupling are neglected. From equation~\eqref{eq:fsde}, the energy density of the effective dark energy component
is given by\begin{equation}
\begin{aligned}\label{FSDEintegrated}
\rho_{\mathrm{FSDE}}(a)
=&
\left(
\rho_{\mathrm{DE},*} a_*^{2\sigma}
-\frac{2\sigma}{3-2\sigma}\rho_{m,0} a_*^{2\sigma-3}
-\frac{2\sigma}{4-2\sigma}\rho_{\gamma,0} a_*^{2\sigma-4}
\right)a^{-2\sigma}
\\[0.5em]
&+
\frac{2\sigma}{3-2\sigma}\rho_{m,0}\,a^{-3}
+
\frac{2\sigma}{4-2\sigma}\rho_{\gamma,0}\,a^{-4}
-2\sigma a^{-2\sigma}
\int_{a_*}^{a}
\rho_{\nu}(\tilde a)\,
\tilde a^{\,2\sigma-1}\,
d\tilde a.
\end{aligned}
\end{equation}
where $a_*$ is a reference scale factor in the early universe, $\rho_{\mathrm{DE},*}$ is the initial dark energy density at that epoch, and $\rho_{m,0}$ and $\rho_{\gamma,0}$ denote the present-day matter and radiation densities, respectively.
 For several neutrino species, the
 integral in equation~\eqref{FSDEintegrated} is  summed over the species.  From equation~\eqref{FSDEintegrated}, it is clear that the radiation- and matter-tracking contributions redshift
faster than the $a^{-2\sigma}$ mode. This slowly evolving mode
(decaying for $\sigma>0$ and growing for $\sigma<0$), eventually dominates at late times.

The last term in equation~\eqref{FSDEintegrated}  is the neutrino contribution to the source-history integral introduced
in Sec. \ref{subsec:constant-log}. We now show that the mass-dependent evolution of neutrinos yields a finite residual
beyond the analytic tracking and boundary terms. This residual is associated with  neutrinos becoming
non-relativistic and shifts the coefficient of the vacuum-like \(a^{-2\sigma}\) mode. 
In general, any cosmological species whose equation of state changes can generate analogous residual contributions to the late-time dark-energy density.

We introduce the dimensionless momentum $y \equiv p/T_{\nu,0}$ and the
scale factor $x \equiv m_\nu \tilde{a}/T_{\nu,0}$. In terms of these variables, the neutrino contribution to dark energy simplifies to:
\begin{align}
\rho_{\rm FSDE}^{(\nu)}(a)={}&-2\sigma a^{-2\sigma}m_\nu^4
\left(\frac{T_{\nu,0}}{m_\nu}\right)^{2\sigma}\frac{g_\nu}{2\pi^2}
\int_{x_*}^{x}dx'\,(x')^{2\sigma-5}
\int_0^\infty dy\,\frac{y^2\sqrt{y^2+(x')^2}}{e^y+1}.
\label{eq:source-nu}
\end{align}
The above reveals an interesting result that the neutrino contribution to the dark energy scales roughly as $m_\nu^4$.
Applying integration by parts three times to the integral
$\mathcal{I}_{\sigma}(x_*, x) = \int_{x_*}^x x'^{\,2\sigma-5} J(x')\,dx'$, where

\[
J(x) \equiv \int_0^\infty
\frac{y^2\sqrt{y^2+x^2}}{e^y+1}\,dy,
\]
allows us to separate the integral into analytical boundary terms and
a residual integral:
\begin{equation}
    \mathcal{I} = \mathcal{B}(x) - \mathcal{B}(x_*) + \mathcal{N}_\sigma(x_*,x).
\end{equation}
where the boundary function and $\mathcal{N}_\sigma(x_*,x)$ are given by:
\begin{equation}
    \mathcal{B}(l) = \frac{l^{2\sigma-4}}{2\sigma-4} J(l) - \frac{l^{2\sigma-3}}{(2\sigma-4)(2\sigma-3)} J'(l) + \frac{l^{2\sigma-2}}{(2\sigma-4)(2\sigma-3)(2\sigma-2)} J''(l),
\end{equation}
\begin{equation}
    \mathcal{N}_\sigma(x_*, x) = \frac{3}{(2\sigma-4)(2\sigma-3)(2\sigma-2)} \int_{x_*}^x x'^{\,2\sigma-1} \left[ \int_0^\infty \frac{y^4}{(y^2+x'^2)^{5/2}(e^y+1)} \, dy \right] dx'.
\end{equation}
Because we typically select an early-universe reference scale of $a_* \sim \mathcal{O}(10^{-9})$, the corresponding dimensionless lower limit is sufficiently small ($x_* \sim \mathcal{O}(10^{-7})$) to justify the following approximation:
\begin{equation}
    J(x_*) \approx \frac{7\pi^4}{120} + \frac{\pi^2}{24}x_*^2, \quad \mathcal{B}(x_* ) \approx \frac{7\pi^4}{120(2\sigma-4)} x_*^{2\sigma-4}-\frac{\pi^2}{48(1-\sigma)}\,x_*^{2\sigma-2}.\label{boundx*}
\end{equation}
The second term in $\mathcal{B}(x_*)$ is the boundary counterpart of the ${\cal O}(m_\nu^2)$ correction to the early neutrino density, $\rho_\nu\simeq\rho_{\gamma,0}^{(\nu)}a^{-4}+\rho_{\nu,2}a^{-2}$ with $\rho_{\nu,2}\equiv g_\nu m_\nu^2T_{\nu,0}^2/48$.
Similarly, if we set the scale factor $a$ near the present epoch
($a=1$), the dimensionless scale factor is typically of order
$x\sim\mathcal{O}(100)$. This value is sufficiently large to justify
  the following approximation:
\begin{equation}
J(x)
\approx
\frac{3}{2}\zeta(3)x
+
\frac{45}{4}\zeta(5)x^{-1}, \quad \mathcal{B}(x)
\approx
\frac{3\zeta(3)}{2(2\sigma-3)}x^{2\sigma-3}.\label{boundx}
\end{equation}
Using equations~\eqref{boundx*} and \eqref{boundx}, the neutrino contribution to the fast-scrambling dark energy density is given as a late-time approximation by:
\begin{align}
    \rho_{\rm FSDE}^{(\nu)}(a) \approx \frac{2\sigma}{3-2\sigma} \rho_{\nu,0} a^{-3}
    - \frac{2\sigma}{4-2\sigma} \rho_{\gamma,0}^{(\nu)} a_*^{2\sigma-4} a^{-2\sigma}
    - \frac{2\sigma}{2-2\sigma} \rho_{\nu,2}\, a_*^{2\sigma-2} a^{-2\sigma}
    + \Delta\rho_{\rm trans} a^{-2\sigma},
    \label{eq:neutrino_fsde_final}
\end{align}
where $\rho_{\nu,0} = [3\zeta(3)/(4\pi^2)] g_\nu m_\nu T_{\nu,0}^3$ is the present-day neutrino  density,
$\rho_{\gamma,0}^{(\nu)} = (7\pi^2/240) g_\nu T_{\nu,0}^4$ denotes the fictitious relativistic radiation energy density the neutrinos would possess if they had remained massless, and $\rho_{\nu,2}$ is the $\mathcal{O}(m_\nu^2)$ coefficient defined below equation~\eqref{boundx*}. 
Thus, we can  write $\rho_{\mathrm{FSDE}}(a)$ as:
\begin{equation}
\begin{aligned}\label{FSDEsimp}
\rho_{\mathrm{FSDE}}(a)
=&
C(a_*)a^{-2\sigma}+
\frac{2\sigma}{3-2\sigma}(\rho_{m,0}+\rho_{\nu,0})\,a^{-3}
+
\frac{2\sigma}{4-2\sigma}\rho_{\gamma,0}\,a^{-4}
+\Delta\rho_{\rm trans}  a^{-2\sigma}.
\end{aligned}
\end{equation}
where
\begin{align*}
C(a_*)={}&\rho_{{\rm DE},*}a_*^{2\sigma}
-\frac{2\sigma}{3-2\sigma}\rho_{m,0}a_*^{2\sigma-3}
-\frac{2\sigma}{4-2\sigma}(\rho_{\gamma,0}+\rho_{\gamma,0}^{(\nu)})a_*^{2\sigma-4}
-\frac{2\sigma}{2-2\sigma}\rho_{\nu,2}a_*^{2\sigma-2},
\end{align*} i.e.\ the coefficient of $a^{-2\sigma}$ that remains once all tracking terms, including the ${\cal O}(m_\nu^2)$ one, are subtracted at $a_*$.
The interesting term is the transition  density term:
\begin{equation}
    \Delta\rho_{\rm trans}  = -\frac{\sigma g_\nu m_\nu^{4-2\sigma}T_{\nu, 0}^{2\sigma}}{\pi^2}\mathcal{N}_{\sigma}(x_*, x).
\end{equation}
By construction, $\Delta\rho_{\rm trans}(a_*)=0$.
Moreover, for massless neutrinos, where $\rho_\nu\propto a^{-4}$,
the residual $\Delta\rho_{\rm trans}(a)$ vanishes identically.
For finite $m_\nu$, $\Delta\rho_{\rm trans}$ evolves and
approaches a finite constant $\Delta C$ once the neutrinos
are sufficiently non-relativistic. It thereby shifts the
late-time coefficient of the $a^{-2\sigma}$ mode from
$C(a_*)$ to $C(a_*)+\Delta C$.
Although $\Delta\rho_{\rm trans}$ saturates once the neutrinos are non-relativistic, most of it is accumulated while they are still relativistic: for $x\ll1$ the inner integral in $\mathcal{N}_\sigma$ behaves as $\int_0^\infty dy\,y^4/[(y^2+x^2)^{5/2}(e^y+1)]\simeq\tfrac12\ln(1/x)$, which is the imprint of the non-analytic ${\cal O}(m_\nu^4\ln a)$ term in $\rho_\nu(a)$ sourced by the low-momentum tail of the distribution. Integrating, one finds
\begin{equation}
\Delta\rho_{\rm trans}\simeq\frac{g_\nu\,\sigma\,m_\nu^4}{32\pi^2}\,L^2\left[1+{\cal O}\!\left(L^{-1},\,\sigma L\right)\right],
\qquad L\equiv\ln\!\left(\frac{T_{\nu,0}}{m_\nu a_*}\right).
\label{eq:trans-asymptotic}
\end{equation}
This approximation agrees with the numerical results to within about \(10\%\) for \(|\sigma|\lesssim10^{-5}\). The error increases to about \(50\%\) at \(\sigma=0.01\), where \(\sigma L\) is no longer sufficiently small. The expression also shows a logarithmic dependence on the reference epoch \(a_*\). 

\begin{figure}[htbp]
    \centering
    \includegraphics[width=0.78\textwidth]{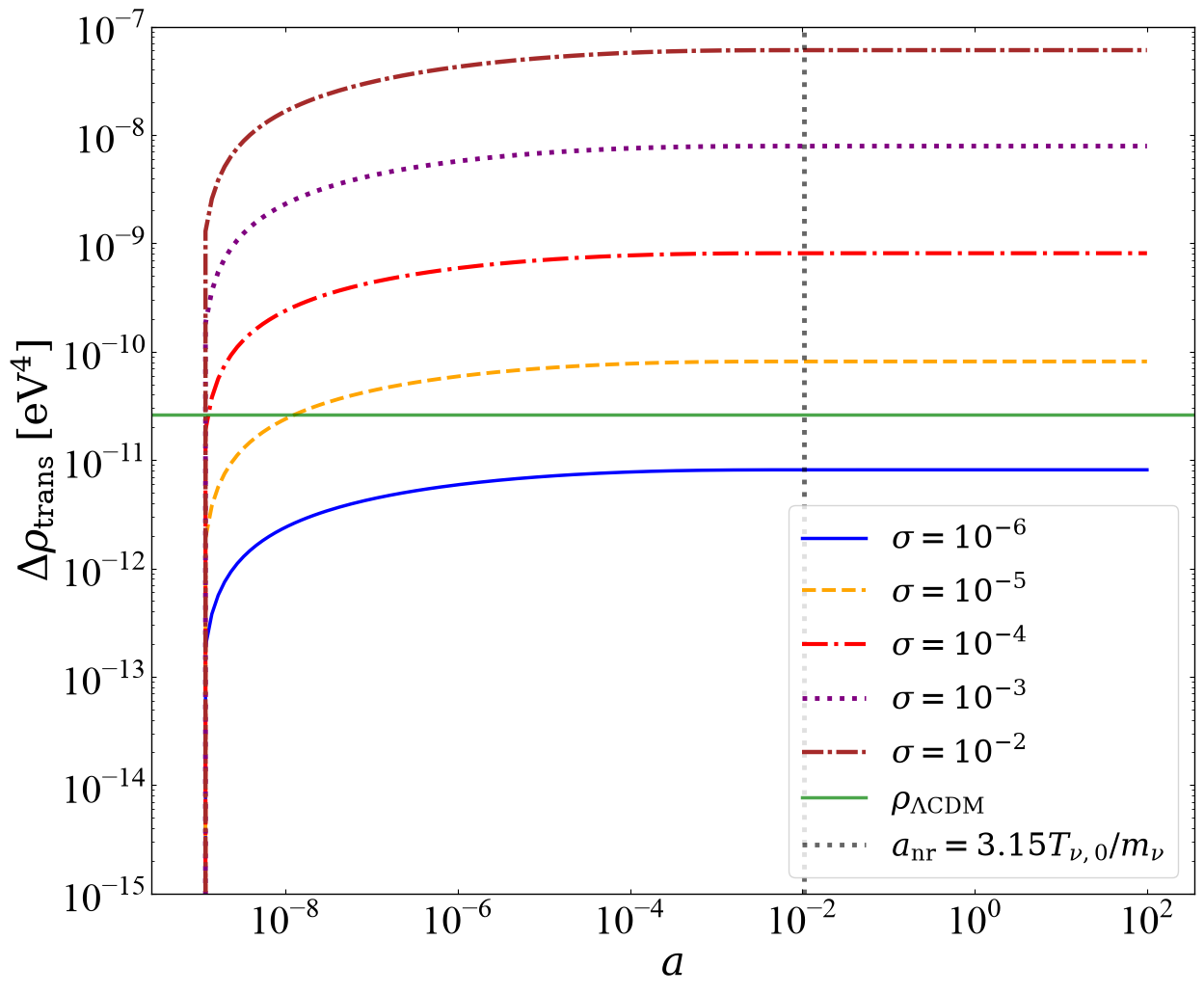}
        \caption{
Evolution of $\Delta\rho_{\rm trans}$ with scale factor $a$ for $\sigma=10^{-6},10^{-5},10^{-4},10^{-3}$ and $10^{-2}$, assuming $m_\nu=0.05\,\mathrm{eV}$ and $g_\nu=2$. The mass-dependent residual vanishes at $a_*=10^{-9}$, accumulates mostly while neutrinos remain relativistic, and approaches a finite late-time plateau. The vertical dotted line marks  the characteristic
non-relativistic transition scale $a_{\rm nr}=3.15T_{\nu,0}/m_\nu$.}
    \label{fig:R_a1}
\end{figure}

Figure~\ref{fig:R_a1} shows the evolution of
the growth and asymptotic saturation of $\Delta\rho_{\rm trans}$ for different $\sigma$.
This saturation indicates that the neutrino transition leaves a finite contribution to the amplitude of the late-time dark-energy density and  this contribution increases with $\sigma$.
 For \(a_*=10^{-9}\), approximately \(95\%\) of the plateau value is reached by \(a\simeq10^{-4}\),  well before the
characteristic non-relativistic transition scale
$a_{\rm nr}\equiv 3.151\,T_{\nu,0}/m_\nu\simeq 0.011$
for $m_\nu=0.05\,\mathrm{eV}$, defined by equating
the mean neutrino momentum to its mass. Interestingly,
for $\sigma\sim10^{-6}$--$10^{-5}$, $\Delta\rho_{\rm trans}$
becomes comparable to the observed dark-energy
density, suggesting that it could
account for  potentially
all of the late-time dark energy if  the initial-condition contribution $C(a_*)$ is negligible, $C(a_*)\approx0$. The initial condition $C(a_*)\approx0$ 
physically  implies that
$\rho_{\rm FSDE}(a_*)$ consists entirely of the
tracking contributions at the reference epoch.

\begin{figure}[htbp]
    \centering
    \includegraphics[width=0.78\linewidth]{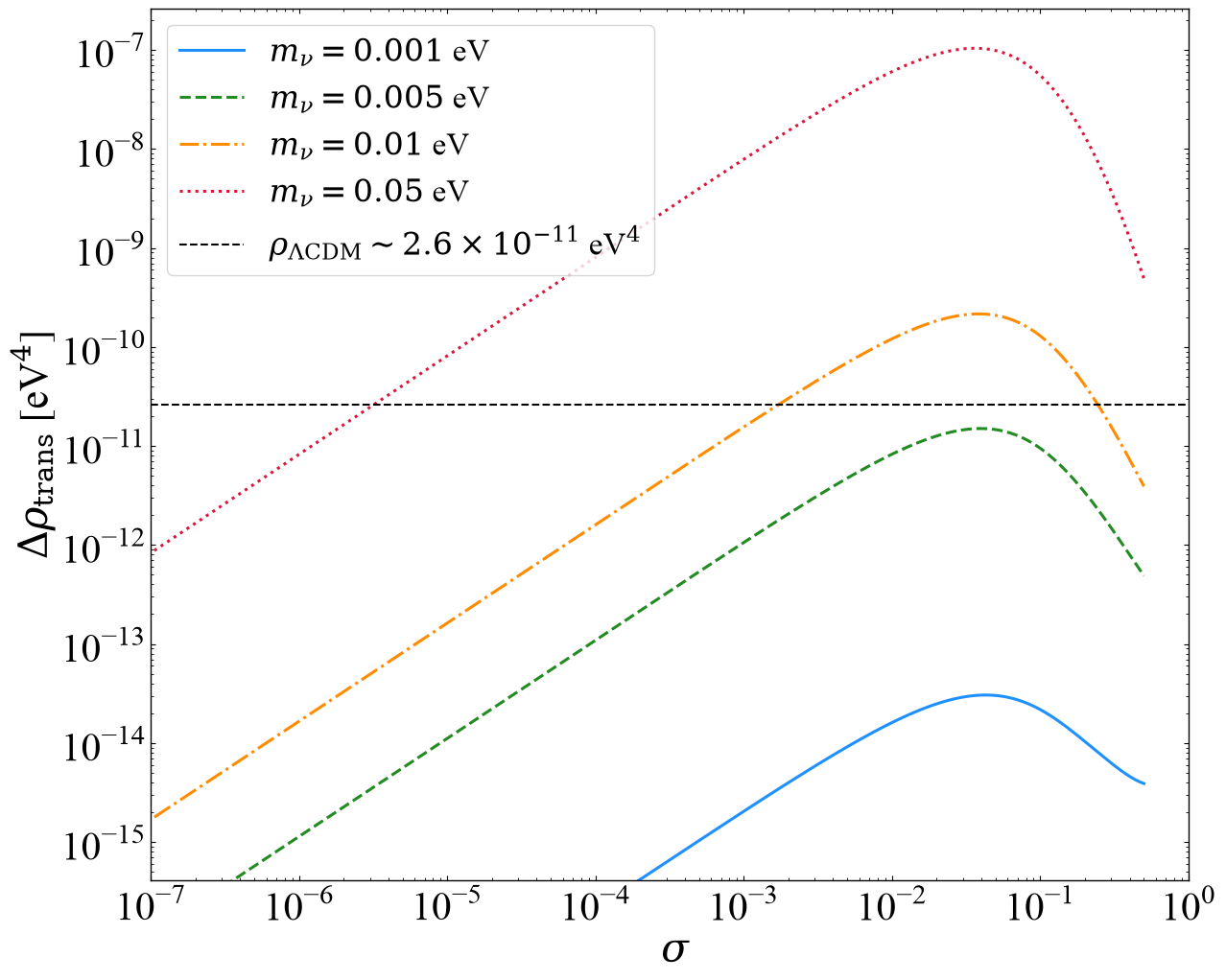}
    \caption{The transition density $\Delta\rho_{\rm trans} $ as a function of the scrambling parameter $\sigma$ for various neutrino masses. The horizontal dashed line indicates the standard $\Lambda$CDM dark energy density.}
    \label{fig:rhotransvssigma}
\end{figure}

Figure~\ref{fig:rhotransvssigma} explores the dependence of the transition density, $\Delta\rho_{\rm trans} $, at the present time, $a=1$, on the scrambling parameter $\sigma$ for different neutrino masses. The amplitude of $\Delta\rho_{\rm trans} $ follows the expected approximate scaling $\Delta\rho_{\rm trans}  \propto m_\nu^4$, as seen in the figure. For small $\sigma$, $\Delta\rho_{\rm trans} $ increases approximately linearly with $\sigma$ (and becomes increasingly  negative as $\sigma$ decreases below 0). For positive $\sigma$, it  reaches a maximum and subsequently decreases as $\sigma$ is increased further.

\begin{figure}[htbp]
    \centering
    \includegraphics[width=0.7\linewidth]{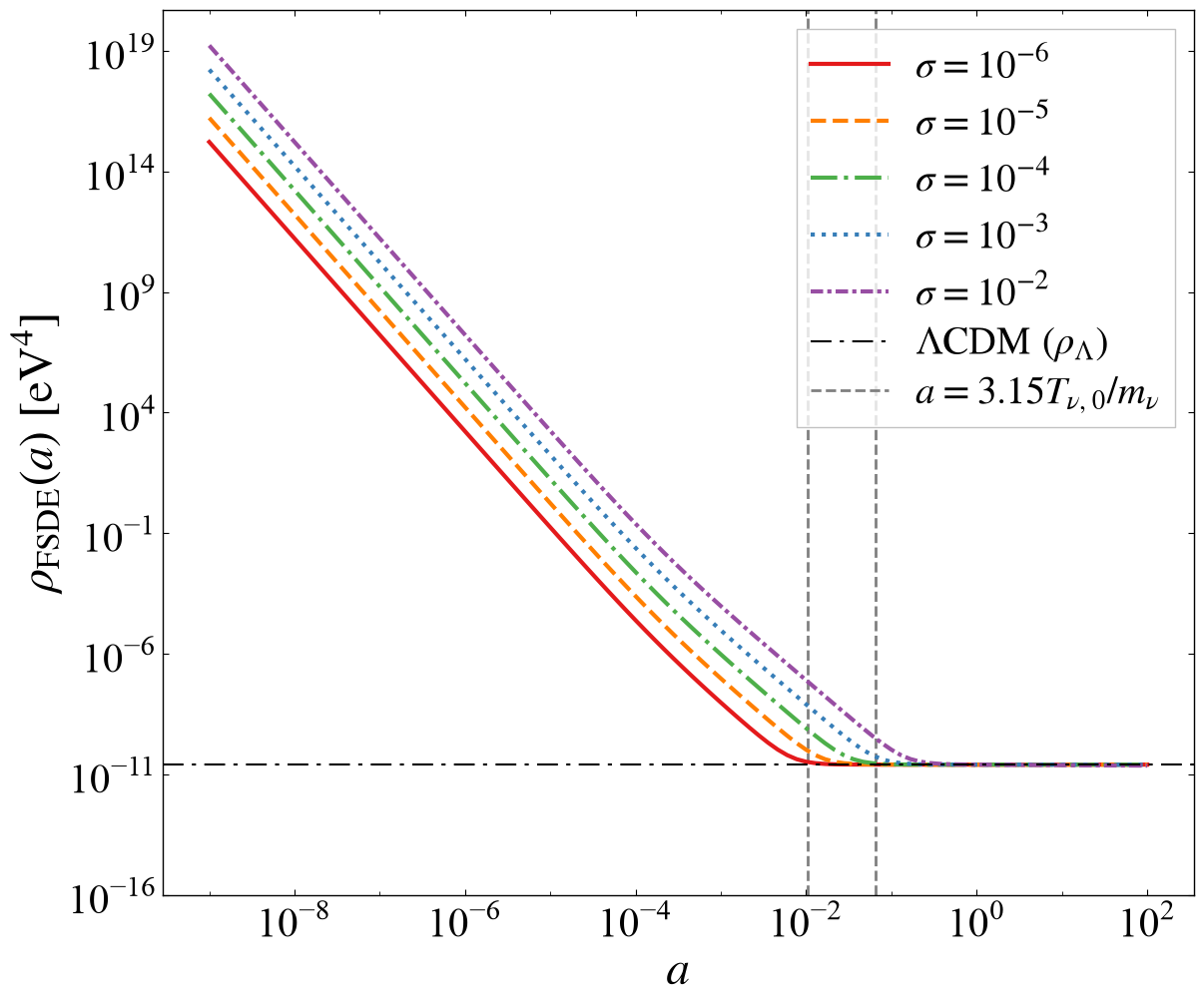}
    \caption{Evolution of $\rho_{\rm FSDE}(a)$ for different values of the
scrambling parameter $\sigma$, held constant along each curve.
We use $h=0.674$, $\Omega_{m,0}=0.315$ for matter excluding
neutrinos, $\Omega_{\gamma,0}=5.38\times10^{-5}$, and
$\Omega_{{\rm DE},0}=0.683$ (fixed by flatness), with the present-day boundary
condition $\rho_{\rm FSDE}(1)=\rho_\Lambda
=\Omega_{{\rm DE},0}\rho_{c,0}$.
The adopted normal-ordering neutrino spectrum has
$m_1=0$, $m_2=0.0086\,\mathrm{eV}$, and
$m_3=0.05\,\mathrm{eV}$.
The vertical dashed lines mark the characteristic
non-relativistic transition scales of the two massive species,
$a_{{\rm nr},i}=3.15T_{\nu,0}/m_i$.
The horizontal dash-dotted line shows the constant
$\Lambda$CDM reference density $\rho_\Lambda$. }
    \label{fig:rho_fsde}
\end{figure}

Figure~\ref{fig:rho_fsde} illustrates the evolution of the effective fast-scrambling dark energy density, $\rho_{\mathrm{FSDE}}$, as a function of the scale factor $a$. This evolution is obtained by 
imposing the present-day dark energy density to be  $\rho_{\rm FSDE}(1)=\rho_\Lambda$ and evolving the background toward earlier epochs.
The ``knee'' where $\rho_{\mathrm{FSDE}}$ breaks away from its matter- and radiation-tracking behavior and begins to behave approximately like $\Lambda$CDM occurs approximately when the matter tracking term falls below the vacuum mode, at $a_{\rm knee}\simeq[2\sigma\rho_{m,0}/(3\rho_\Lambda)]^{1/3}\simeq0.007$--$0.014$ for $\sigma=10^{-6}$--$10^{-5}$ with the present-day boundary condition $\rho_{\rm FSDE}(1)=\rho_\Lambda$ adopted here.  This is close to the characteristic non-relativistic
transition scale, $a_{\rm nr}\simeq0.011$, of the
heaviest neutrino, $m_3\simeq0.05\,\mathrm{eV}$,
in the  normal-ordering spectrum under consideration,
assuming the lightest neutrino has $m_1=0$. However,  its proximity to the neutrino transition does not, by itself, establish that the transition causes the knee.

\begin{figure}[htbp]
    \centering
    \includegraphics[width=0.78\linewidth]{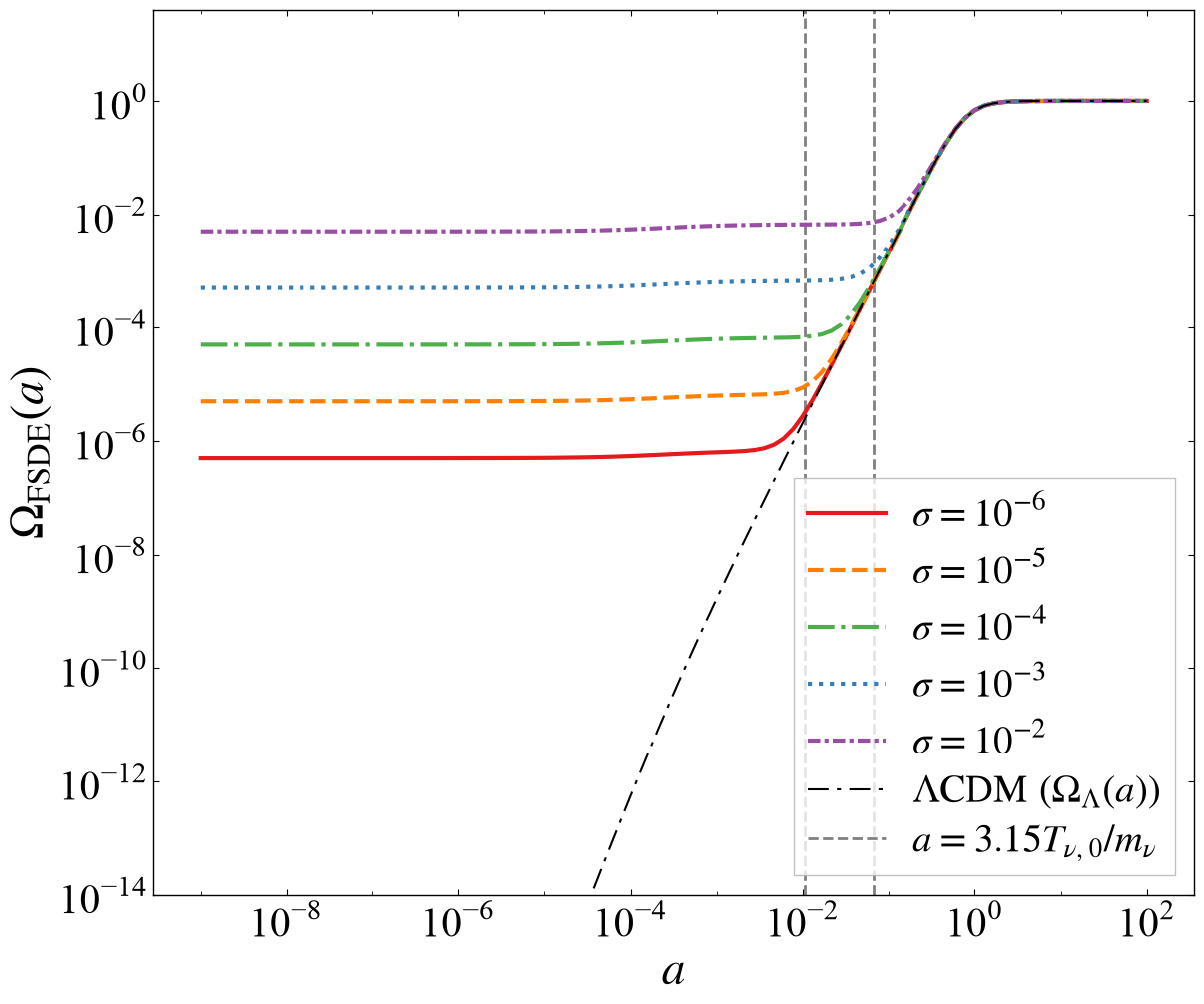}
    \caption{Evolution of the fractional fast-scrambling dark energy density, $\Omega_{\mathrm{FSDE}}$, with the scale factor $a$ for different scrambling parameters $\sigma$. The cosmological parameters are same as used in Figure~\ref{fig:rho_fsde}. The vertical dashed lines indicate the epochs at which the neutrino species become non-relativistic.  The $\Lambda$CDM evolution is shown for comparison.}
    \label{fig:Omega_fsde}
\end{figure}

Figure~\ref{fig:Omega_fsde} shows the evolution of the fractional energy density, $\Omega_{\mathrm{FSDE}}$, with the scale factor $a$. During radiation domination, it approaches a constant tracking value, $\Omega_{\mathrm{FSDE}}\simeq\sigma/2$, consistent with equation~\eqref{FSDEsimp}, and then shifts to a slightly higher matter-tracking plateau  at $\Omega_{\mathrm{FSDE}}\simeq2\sigma/3$. As the slowly evolving vacuum-like mode becomes
dominant, the FSDE fraction rises rapidly,
closely following the standard $\Lambda$CDM
evolution.

Two remarks are in order. First, 
the parameter range $\sigma\sim10^{-6}$--$10^{-5}$
corresponds to
$\Gamma=\sigma\ln(M_{\rm Pl}^2/H^2)\sim10^{-3}$,
well below the $O(1)$ magnitude motivated by the
fast-scrambling timescale and the central proxy
estimates $\Gamma\simeq-3$ of
Sec.~\ref{sec:sigma_bounds}. For $|\sigma|\sim0.01$ the neutrino residual is several thousand times larger than $\rho_\Lambda$ (and negative for $\sigma<0$), so that it must be compensated by the early-time constant $C(a_*)$: it then becomes part of the cosmological-constant problem rather than its solution. Second, by equation~\eqref{eq:trans-asymptotic} the value of $\sigma$ for which $\Delta\rho_{\rm trans}=\rho_\Lambda$ depends on $a_*$; the coincidence $\Delta\rho_{\rm trans}\simeq\rho_\Lambda$ for $\sigma\sim10^{-6}$--$10^{-5}$ is a restatement of the well-known numerical coincidence $m_\nu^4\sim10^{5}\rho_\Lambda$, with the small factor supplied by $\sigma\ln^2(T_{\nu,0}/m_\nu a_*)$. Thus, while the mass-dependent neutrino history
provides a mechanism for generating a finite
late-time dark-energy contribution, its ability
to reproduce the observed abundance depends
on the scrambling parameter and the
initial conditions.

\section{Discussion and outlook}
\label{sec:discussion}

In this work, we have investigated a phenomenological modification of the cosmological background equations motivated by the de Sitter scrambling timescale. The additional term in the Raychaudhuri equation destabilizes the exact de Sitter solution for either sign of its coefficient $\Gamma$, leading to a slow decay of the Hubble parameter for $\Gamma>0$ and a slow growth for $\Gamma<0$, both on a characteristic timescale $\sim\fs/|\Gamma|$. In the slowly varying logarithm approximation, the model admits an analytic description in terms of a vacuum-like mode and a contribution sourced by the history of the cosmological fluids. The vacuum mode scales as $a^{-2\sigma}$ and has an effective equation of state $w_{\rm vac}=-1+2\sigma/3$, exhibiting phantom-like evolution for $\sigma<0$ and quintessence-like evolution for $\sigma>0$. The source-history contribution, on the other hand, generates terms that track the dominant matter and radiation densities, as well as residual contributions associated with the change in the equation of state of the cosmological species.

The TCC and fast-scrambling arguments motivate the characteristic timescale of the de Sitter instability, but do not determine its direction.  For $\Gamma<0$, the Hubble parameter grows toward the Planck scale, which is reached after approximately $100$ Hubble times for $\sigma_0\simeq-0.01$. This corresponds to a slow Big Rip evolution in the constant-$\sigma$ approximation, although the exact solution reaches the Planck scale without developing the corresponding divergence.  
Furthermore, neither sign of $\Gamma$ satisfies the integrated TCC bound by itself (Sec.~\ref{subsec:vacuum}). The model instead realizes the scrambling timescale as the rate at which de Sitter space is deformed, and an exit from accelerated expansion must come from physics beyond the effective description. Whether such a slow Big Rip is compatible with the microscopic picture of horizon scrambling is an open question.

By mapping the early-time tracking terms of the model onto a cosmological gravitational coupling, we derived proxy estimates of the scrambling parameter from BBN abundances~\cite{Yeh:2026pil}, from BBN bounds on $G$~\cite{Alvey:2019ctk}, and from the recent CMB+BAO+SN fit of the cosmic-glitch model~\cite{Wen:2026xlj}: $\Gamma=-3.7\pm6.1$, $-3.9\pm10.8$ and $-2.8\pm0.9$, respectively ($68\%$ CL). The estimates are consistent with the ${\cal O}(1)$ magnitude anticipated from the scrambling argument, and the tightest translates into a negative value at the $3\sigma$ level. As no likelihood analysis of the fast-scrambling model itself has been performed, these numbers should be regarded as illustrative.

We compared the predicted dark-energy evolution with the DESI DR2 CPL results in Sec.~\ref{sec:DESIoverlay}. For the fixed values of fast-scrambling parameters considered, the negative-$\sigma$ branch lies within the CPL posterior bands for $z\gtrsim1$. This comparison shows that the fast-scrambling model can exhibit late-time density evolution qualitatively similar to that inferred from DESI.

A further consequence of our fast-scrambling model is that  changes in the equation of state of cosmological species can leave a residual imprint on the late-time dark-energy density. We have explored this mechanism for  neutrinos, whose transition from relativistic to non-relativistic behavior generates a finite late-time contribution to the effective dark-energy density. In Sec.~\ref{sec:neutrino}, we showed that this transition's contribution to dark energy density scales approximately as $m_\nu^4$ and approaches a constant once the neutrinos become non-relativistic. 
For $\sigma\sim10^{-6}$--$10^{-5}$, the neutrino contribution
can become comparable to the observed dark-energy density.
It could account for essentially the entire late-time amplitude
if $C(a_*)$ is negligible, corresponding to an initial
dark-energy density set by its early-time tracking value.  This possibility, however, requires $\Gamma\sim {\cal{O}}(10^{-3})$, considerably smaller than the ${\cal O}(1)$  magnitude motivated by the scrambling argument and the proxy estimates.

An important  question is whether the phenomenological background evolution admits a consistent covariant realization. The no-go proposition in Sec.~\ref{subsec:nogo} shows that the model cannot arise from a local minisuperspace action of the
$f(R,H)$ class. However, it does not exclude more general covariant constructions and in Sec.~\ref{subsec:moregeneralactions}, we present a scalar-clock construction that reproduces the constant $\sigma$ background equation, but whose scalar mode has $c_s^2<0$ at all epochs (Appendix~\ref{app:perturbations}).

An alternative description of the same background evolution can be 
obtained by interpreting the fast-scrambling contribution as an independently conserved dark-energy fluid with equation of state
\begin{equation}
P_D=-\rho_D+\frac{2}{9}\sigma\Theta_D^2,
\qquad
\nabla_\mu T_D^{\mu\nu}=0,
\end{equation}
where $\Theta_D\equiv\nabla_\mu u_D^\mu$. In a comoving FLRW background,
$\Theta_D=3H$, and hence $\rho_D+P_D=2\sigma H^2$. The Einstein equations
therefore reproduce
\begin{equation}
\dot H=-\frac{1}{2}(\rho+P)-\frac{1}{2}(\rho_D+P_D)
=-\frac{1}{2}(\rho+P)-\sigma H^2.
\end{equation}
The dark-energy density evolves as $\dot{\rho}_D=-6\sigma H^3$, recovering
the $\Lambda$CDM limit as $\sigma\to0$. In this form the model is a bulk-viscous dark-energy fluid with an expansion-dependent viscosity, $P_D=-\rho_D-\zeta\,\Theta_D$ with $\zeta=-\tfrac{2}{9}\sigma\Theta_D$ (positive for the phantom branch $\sigma<0$). First-order, Eckart-type viscous fluids are known to suffer from causality and stability problems~\cite{Hiscock:1985zz}, consistent with the gradient instability of the clock construction found in Appendix~\ref{app:perturbations}; a consistent treatment of perturbations will require either a second-order (Israel--Stewart-type) completion or additional operators in the field-theoretic description. We defer the full
theoretical development of this covariant framework, including its
perturbations and stability analysis, to a forthcoming companion paper.
We discuss two possible future directions. Firstly, one could construct a perturbatively stable completion. The background fixes only $\rho_D+p_D=2\sigma H^2$, while the kineticity and braiding functions of the effective theory of dark energy are free. 
The minimal clock construction, with $\alpha_B=4\sigma/3$, is unstable; a stable realization of the $\sigma<0$ branch requires a braiding function $\alpha_B(t)$ of suitable sign and magnitude, as for other phantom-like backgrounds in kinetic gravity braiding~\cite{Deffayet:2010qz}.
A particularly simple extension is to add
the operator $\tfrac{1-\lambda}{2}\Theta^2$
to~\eqref{eq:clock-lagrangian}.
This contributes $(\lambda-1)\dot\Theta$
to $\rho_D+p_D$ and modifies the gravitational 
kinetic term in unitary gauge to
$K_{ij}K^{ij}-\lambda K^2$.
Here $\lambda$ is the parameter of
Ho\v{r}ava--Lifshitz (HL) gravity
~\cite{Horava:2009uw}, with $\lambda=1$
corresponding to general relativity. On its own ($\sigma=0$),  this operator is associated
with the low-energy limit of non-projectable
HL gravity, equivalently the quadratic
cuscuton~\cite{Afshordi:2009tt}. It rescales the gravitational coupling felt by the expansion, $G_{\rm N}/G_{\rm cosmo}=\tfrac12(3\lambda-1)$, without adding a propagating degree of freedom, and is therefore precisely the HL/cuscuton realization of the cosmic glitch of Sec.~\ref{sec:sigma_bounds}, with a constant $\Omega_g=-\tfrac32(\lambda-1)$~\cite{Wen:2023wes}. For $\sigma\neq0$ the scalar mode becomes dynamical, with $\mathcal A\simeq-\sigma$ and a Lifshitz-like dispersion relation in the dark-energy era, $\omega^2\simeq-\tfrac53\,k^2/a^2-\tfrac{\lambda-1}{2\sigma}\,k^4/(a^4H^2)$ (Appendix~\ref{app:perturbations}). On the ghost-free branch $\sigma<0$, all modes with $k/aH>\sqrt{10|\sigma|/3(\lambda-1)}$ are therefore stable for $\lambda>1$, which is also the sign favored by the cosmic-glitch data ($\Omega_g<0$). The remaining negative $c_s^2$ at horizon scales is not a genuine
instability, as it is tied to the Hubble scale.
In particular, a comoving mode spends only of order a Hubble time
in this band during decelerated expansion.
Similarly, in the dark-energy era, its total growth after entering
the band is bounded by
$\exp\!\big[\sqrt{50|\sigma|/[9(\lambda-1)]}\big]$,
which is $\simeq10$ for $\lambda-1=|\sigma|$
and $\simeq2$ for $\lambda-1=10|\sigma|$.
Thus, in both cases, the growth remains controlled
rather than exponential.
 The operator therefore regulates the short-wavelength instability of the vacuum-era construction. It does not, however, by itself establish cosmological viability, which requires the matter sector in the quadratic action, the transitions between eras and the strong-coupling scale to be examined; nor does it help the $\sigma>0$ branch of Sec.~\ref{sec:neutrino}, for which $\mathcal A\simeq-\sigma<0$ remains a ghost. In this extension the effective glitch is $\Omega_g\simeq-\tfrac32(\lambda-1)+\tfrac23\sigma$ in the matter era, so that the CGG constraint of Sec.~\ref{sec:sigma_bounds} bounds this combination rather than $\sigma$ alone; a full analysis, including the matter era, is left to future work.

Secondly, we need to perform a Markov chain Monte Carlo analysis using the latest DESI BAO,
CMB, and Type Ia supernova data to constrain $\sigma$ and the
relevant cosmological parameters, and assess the fit of the
fast-scrambling model relative to $\Lambda$CDM and the CPL model. Fast scrambling cosmology thus provides a concrete bridge between the trans-Planckian censorship and fast-scrambling conjectures and late-time observational cosmology. This framework probes the same horizon-scale physics as the search for black hole echoes~\cite{Abedi:2016hgu,OshitaAfshordi2023}, but from the opposite end: the late-time expansion in place of the ringdown.

\acknowledgments
We thank Ohoud Alharbi, Robert Brandenberger, and Cumrun Vafa for helpful discussions. NA acknowledges support from the Natural Sciences and Engineering Research Council of Canada (NSERC) and Perimeter Institute for Theoretical Physics. Research at Perimeter Institute is supported in part by the Government of Canada through the Department of Innovation, Science and Economic Development Canada and by the Province of Ontario through the Ministry of Colleges and Universities. 
AS acknowledges support from the Kishore Vaigyanik
Protsahan Yojana (KVPY), Department of Science and
Technology (DST), Government of India. 

\section*{Use of AI tools}
Large language models (Anthropic's Claude and OpenAI's ChatGPT) were used to assist with editing the text, with literature searches, and with the symbolic computations behind Appendix~\ref{app:perturbations}; all derivations and results were checked and reproduced by the authors, who take full responsibility for the content.

\section*{Data availability}
No new observational data were generated in this work. The DESI DR2 posterior chains used in Sec.~\ref{sec:DESIoverlay} are publicly available at \url{https://data.desi.lbl.gov/public/papers/y3/bao-cosmo-params/}, and the scripts used to produce the figures are available from the corresponding author upon request.

\appendix
\section{Linear perturbations of the clock construction}
\label{app:perturbations}

In the unitary gauge $\phi=t$, we have $-X=N^{-2}$ and $\Theta=K$ where $K$ is the trace of the extrinsic curvature of the constant-time hypersurfaces. Therefore, the Lagrangian~\eqref{eq:clock-lagrangian} becomes 
\begin{equation}
    F(X=-N^{-2}, \Theta=K, \phi=t)=-V(t)-\tfrac{2\sigma}{9}K^2\ln N,
\end{equation} and the action~\eqref{eq:clock-action} takes the form
\begin{equation}
 S=\int d^4x\,N\sqrt{h}\left[\frac{1}{2}\left(K_{ij}K^{ij}-K^2+R^{(3)}\right)+F(N,K,t)\right],
\end{equation}
a spatially covariant theory of the type covered by the effective theory of dark energy~\cite{Gleyzes:2014rba}. Here we have set $\mpl=1$ and omitted $S_{\rm other}$. 

In unitary gauge, without loss of generality, the scalar modes can be described by the following perturbations \cite{Gleyzes:2014rba, Maldacena:2002vr}:
\begin{equation}N=1+\delta N, \quad h_{ij}=a^2e^{2\zeta}\delta_{ij},\quad  N^i
= \delta^{ij} \partial_j \psi.\end{equation}
Expanding the action to first order in the
perturbations yields the background equations
of motion,
$3H^2=V+2\sigma H^2$ and
$2\dot H+3H^2=V$.
 These fix $V=(3-2\sigma)H^2$ and reproduce $\dot H=-\sigma H^2$, which we use
in the subsequent quadratic expansion. In Fourier space, the quadratic order terms in the perturbations are
\begin{align}
 \mathcal L_2={}&a^3\Big[-3\dot\zeta^2+2(3-2\sigma)H\,\delta N\,\dot\zeta-3(1-\sigma)H^2\delta N^2-18H\zeta\dot\zeta-9(3-\sigma)H^2\zeta^2\Big]\nonumber\\
 &+a\,k^2\Big[\zeta^2+2\,\delta N\,\zeta-2\,\psi\,\dot\zeta+2\big(1-\tfrac23\sigma\big)H\,\delta N\,\psi\Big].
 \label{eq:app-L2}
\end{align}
Variation of the Lagrangian with respect to the shift $\psi$ and the lapse $\delta N$ yields two constraints which on substituting back into \eqref{eq:app-L2} gives us the  following up to boundary terms
\begin{equation}
 \mathcal{L}_2=a^3\,\mathcal A(\sigma)\left[\dot\zeta^2-c_s^2(\sigma)\frac{k^2}{a^2}\zeta^2\right],
 \qquad
 \mathcal A(\sigma)=-\frac{3\sigma(3-4\sigma)}{(3-2\sigma)^2},
 \qquad
 c_s^2(\sigma)=-\frac{5(3-2\sigma)}{3(3-4\sigma)}.
 \label{eq:app-cs2}
\end{equation}
For $|\sigma|\ll1$, we have that $\mathcal A(\sigma)\simeq-\sigma$ and $c_s^2(\sigma)\simeq-5/3$. The tensor sector is unaffected, since $F(N, K, t)$ depends on $K_{ij}$ only through its trace. Comparing~\eqref{eq:app-L2} with the general quadratic action of refs.~\cite{Gleyzes:2014rba,Bellini:2014fua} identifies the parameters~\eqref{eq:alphas}, with $\alpha_B$ in the convention of ref.~\cite{Bellini:2014fua}. The standard expression for the sound speed in the presence of minimally coupled matter~\cite{Bellini:2014fua} then reproduces~\eqref{eq:app-cs2} for $\rho=0$ and gives equation~\eqref{eq:cs2} in general. As a check of the procedure, replacing $F$ by the quintessence Lagrangian and $P(X)=X^2$ $k$-essence Lagrangian returns $c_s^2=1$ and $c_s^2=1/3$, respectively.

The same procedure applies to the extended Lagrangian $F=-V(\phi)+\tfrac{\sigma}{9}\Theta^2\ln(-X)+\tfrac{1-\lambda}{2}\Theta^2$ of Sec.~\ref{sec:discussion}, whose last term contributes $(\lambda-1)\dot\Theta$ to $\rho_D+p_D$ and, in unitary gauge, turns the gravitational kinetic term into $K_{ij}K^{ij}-\lambda K^2$. The background equations become $V=\big[\tfrac32(3\lambda-1)-2\sigma\big]H^2$ and $\tfrac12(3\lambda-1)\dot H=-\sigma H^2$. Since $K^2$ contains $(\partial^2\psi)^2$, the shift $\psi$ is no longer a Lagrange multiplier, but it can still be eliminated algebraically in Fourier space together with $\delta N$. For the vacuum background, one finds 
\begin{align}
 \mathcal A&=-\frac{3\sigma\,[3(3\lambda-1)-8\sigma]}{9(3\lambda-1)-3(3\lambda+5)\sigma+8\sigma^2},\nonumber\\
 c_s^2(k)&=-\frac{9(\lambda-1)(3\lambda-1)\,k^2/(a^2H^2)+\sigma\left[3(9\lambda^2+18\lambda-7)-8\sigma(3\lambda+2)\right]}{3\sigma(3\lambda-1)[3(3\lambda-1)-8\sigma]},
 \label{eq:app-lambda}
\end{align}
which reduces to~\eqref{eq:app-cs2} for $\lambda=1$. For $\sigma=0$, the kinetic coefficient vanishes identically and the scalar mode is non-dynamical, as expected for the low-energy limit of non-projectable Ho\v{r}ava--Lifshitz gravity~\cite{Afshordi:2009tt}, while for $|\sigma|,|\lambda-1|\ll1$ one has $\mathcal A\simeq-\sigma$ and $c_s^2\simeq-\tfrac53-\tfrac{\lambda-1}{2\sigma}\,k^2/(a^2H^2)$, which is the dispersion relation quoted in Sec.~\ref{sec:discussion}.

\bibliography{reference}
\bibliographystyle{JHEP}

\end{document}